\documentclass[12pt]{article}

\makeatletter
\renewcommand\section{\@startsection {section}{1}{\z@}%
                                   {-3.5ex \@plus -1ex \@minus -.2ex}
                                   {2.3ex \@plus.2ex}%
                                   {\normalfont\large\bfseries}}
\renewcommand\subsection{\@startsection{subsection}{2}{\z@}%
                                     {-3.25ex\@plus -1ex \@minus -.2ex}%
                                     {1.5ex \@plus .2ex}%
                                     {\normalfont\bfseries}}
\makeatother

\def\IZ{\relax\ifmmode\mathchoice
{\hbox{\cmss Z\kern-.4em Z}}{\hbox{\cmss Z\kern-.4em Z}}
{\lower.9pt\hbox{\cmsss Z\kern-.4em Z}} {\lower1.2pt\hbox{\cmsss
Z\kern-.4em Z}}\else{\cmss Z\kern-.4em Z}\fi}
\def\IR{\relax{\rm I\kern-.18em R}}

\def\one{{\hbox{ 1\kern-.8mm l}}}

\newlength{\bredde}
\def\slash#1{\settowidth{\bredde}{$#1$}\ifmmode\,\raisebox{.15ex}{/}
\hspace*{-\bredde} #1\else$\,\raisebox{.15ex}{/}\hspace*{-\bredde}
#1$\fi}

\newsavebox{\zzzbar}
\sbox{\zzzbar}
  {\setlength{\unitlength}{0.9em}
  \begin{picture}(0.6,0.7)
  \thinlines
  \put(0,0){\line(1,0){0.6}}
  \put(0,0.75){\line(1,0){0.575}}
  \multiput(0,0)(0.0125,0.025){30}{\rule{0.3pt}{0.3pt}}
  \multiput(0.2,0)(0.0125,0.025){30}{\rule{0.3pt}{0.3pt}}
  \put(0,0.75){\line(0,-1){0.15}}
  \put(0.015,0.75){\line(0,-1){0.1}}
  \put(0.03,0.75){\line(0,-1){0.075}}
  \put(0.045,0.75){\line(0,-1){0.05}}
  \put(0.05,0.75){\line(0,-1){0.025}}
  \put(0.6,0){\line(0,1){0.15}}
  \put(0.585,0){\line(0,1){0.1}}
  \put(0.57,0){\line(0,1){0.075}}
  \put(0.555,0){\line(0,1){0.05}}
  \put(0.55,0){\line(0,1){0.025}}
  \end{picture}}

\newcommand{\ena}{\end{eqnarray}}
\newcommand{\beqa}{\begin{eqnarray}}
\newcommand{\eeqa}{\end{eqnarray}}
\newcommand{\bea}{\begin{eqnarray}}
\newcommand{\eea}{\end{eqnarray}}

\newcommand{\be}{\begin{equation}}
\newcommand{\ee}{\end{equation}}

\usepackage{graphicx}

\newcommand{\beq}{\begin{equation}}
\newcommand{\eeq}{\end{equation}}
\newcommand{\ber}{\begin{array}}
\newcommand{\eer}{\end{array}}

\newcommand{\del}{\partial}

\newcommand{\dsty}{\displaystyle}

\newcommand{\de}{\delta}

\newcommand{\ga}{\gamma}

\usepackage{amsmath}
\usepackage{amssymb}

\begin{document}
\begin{titlepage}
\begin{flushright}
arXiv:0905.1843
\end{flushright}
\vfill
\begin{center}
{\LARGE\bf p-branes on the waves}    \\
\vskip 10mm
{\large Ben Craps,$^a$ Frederik De Roo$^{a,b,}$\footnote{Aspirant FWO}, Oleg Evnin$^a$ and Federico Galli$^a$}
\vskip 7mm
{\em $^a$ Theoretische Natuurkunde, Vrije Universiteit Brussel and\\
The International Solvay Institutes\\ Pleinlaan 2, B-1050 Brussels, Belgium}
\vskip 3mm
{\em $^b$ Universiteit Gent, IR08\\Sint-Pietersnieuwstraat 41, B-9000 Ghent, Belgium}
\vskip 3mm
{\small\noindent  {\tt Ben.Craps@vub.ac.be, fderoo@tena4.vub.ac.be, eoe@tena4.vub.ac.be, fgalli@tena4.vub.ac.be}}
\end{center}
\vfill

\begin{center}
{\bf ABSTRACT}\vspace{3mm}
\end{center}

We present a large family of simple, explicit ten-dimensional supergravity solutions describing extended extremal supersymmetric Ramond-Ramond $p$-branes embedded into time-dependent dilaton-gravity plane waves of an arbitrary (isotropic) profile, with the brane world-volume aligned parallel to the propagation direction of the wave. Generalizations to the non-extremal case are not analyzed explicitly, but can be pursued as indicated.
\vfill

\end{titlepage}

\section{Introduction}

Dilaton-gravity plane waves play a special role in string theory and related approaches to quantum gravity, since they provide a rare example of tractable strongly curved (possibly singular) space-time backgrounds that depend on
(light-cone) time (see, for instance, \cite{HorowitzSteif,PRT}, as well as the recent work by some of the present authors \cite{self}). Furthermore they permit a formulation of (time-dependent) matrix theories of quantum gravity \cite{MBB,ppmatrix}.

A likewise prominent role is accorded to the $p$-brane supergravity solutions (see, e.g., \cite{stelle}). Through their connection with the D-branes of string theory, they lead to the formulation
of the AdS/CFT correspondence \cite{AdSreview} and its generalizations to different dimensions \cite{IMSY}.

Hence, it appears important to derive supergravity solutions
describing $p$-branes embedded into dilaton-gravity plane waves. The simplest of these
solutions are supersymmetric configurations corresponding to extremal $p$-branes aligned
along the propagation direction of the plane wave (the existence of such configurations
can be suggested by the DBI worldvolume analysis for the corresponding D-branes).
Some considerations of these, and related, configurations have been undertaken
in \cite{mbbsolns, intersect, intersect2,blackstring} for highly specific choices of the plane wave profile  (for some other related publications, see \cite{LozanoTellechea:2002pn,Bain:2002nq,Mas:2003uk,Biswas:2002yz}). Our present purpose is to derive this type of solutions without any assumptions regarding the functional shape of the asymptotic plane wave.

\section{Some general considerations}

We shall start by inspecting the ten-dimensional Einstein-frame supergravity equations of motion (see, e.g., \cite{stelle}):
\begin{eqnarray}
&\dsty R_{\mu\nu} = \frac12 \del_\mu
\phi\, \del_\nu \phi + \sum_{N} \frac{1}{2 n_N!}
 e^{a_N \phi} \left[ n_N \left( F_{n_N}^2 \right)_{\mu\nu}
 - \frac{n_N -1}{8} F_{n_N}^2 g_{\mu\nu} \right],
\label{Einstein}\\
&\dsty \Box \phi = \sum_{N} \frac{a_N}{2 n_N!}
e^{a_N \phi} F_{n_N}^2,
\label{dila}\\
&\dsty \del_{\mu_1} \left(
\sqrt{- g} e^{a_N \phi} F^{\mu_1 \cdots \mu_{n_N}} \right) = 0,
\label{field}\\
&\dsty \del_{[\mu} F_{\mu_1 \cdots \mu_{n_N}]} =0,
\label{bianchi}
\end{eqnarray}
where $N$ labels the various form fields of the theory, with field strengths $F_{n_N}$ of rank $n_N$, and $a_N = (5-n_N)/2$ for Ramond-Ramond form fields, for the following ansatz:
\begin{align}
&ds_E^2=A(u,r)\left(-2du dv + K(u,r)du^2+dy_\alpha^2\right) + B(u,r)\,dx_a^2,\label{metricansatz}\\
&\phi=\phi(u,r),\\
&F_{uv\alpha_1\cdots\alpha_{p-1}a}=\frac{x^a}{r}\,\frac{F(u,r)A^{(p+1)/2}e^{\frac{p-3}2\phi}}{B^{(7-p)/2}}\epsilon_{\alpha_1\cdots\alpha_{p-1}}\left[\frac{1}{\sqrt{2}}\right]_{p=3}\qquad (p\le 3),\\
&F_{a_1\cdots a_{8-p}}=\frac{x^a}{r}\,F(u,r)\epsilon_{a_1\cdots a_{8-p}a}\left[\frac{1}{\sqrt{2}}\right]_{p=3}\qquad (p\ge 3),\label{mformansatz}
\end{align}
where $r^2=x^ax^a$, $p$ is the number of spatial dimensions of the $p$-brane, $F$ is the field strength of the corresponding Ramond-Ramond form, $\alpha$ runs from 1 to $p-1$ and $a$ runs from 1 to $9-p$; the factors of $1/\sqrt{2}$ are only inserted into the form field ansatz for the self-dual case $p=3$. This ansatz is not the most general one allowed by the symmetries (in particular, there is no Poincar\'e symmetry relating $g_{uv}$ and $g_{\alpha\alpha}$ when the $u$-dependences are non-trivial), however it will prove sufficiently general for our purposes.

We shall refer the reader to the appendix for the explicit ``raw'' form of the equations of motion for our ansatz, and only present here their convenient combinations. Throughout, prime denotes derivatives with respect to $r$ and dot denotes derivatives with respect to $u$. First of all, the equations
for the form (\ref{field}-\ref{bianchi}) can be integrated straightforwardly to yield
\be
F(u,r)=\frac{Q}{r^{8-p}},
\label{formsolved}
\ee
where $Q$ measures the brane charge. With these dependences, the $uv$-component of Einstein's equations (identical to the $\alpha\alpha$-components) can be written as:
\be
\left(r^{8-p}A^{(p+1)/2}B^{(7-p)/2}\frac{A'}A\right)'=\frac{7-p}8\hspace{1pt}Q^2\hspace{1pt}\frac{e^{\frac{p-3}2\phi}A^{(p+1)/2}}{r^{8-p}B^{(7-p)/2}}.
\label{uvform}
\ee
The dilaton equation (\ref{dila}) gives
\be
\left(r^{8-p}A^{(p+1)/2}B^{(7-p)/2}\phi'\right)'=\frac{p-3}4\hspace{1pt}Q^2\hspace{1pt}\frac{e^{\frac{p-3}2\phi}A^{(p+1)/2}}{r^{8-p}B^{(7-p)/2}}.
\label{dilatonform}
\ee
The $ab$-components of Einstein's equations yield (from terms proportional to $\de_{ab}$)
\be
\left(r^{8-p}A^{(p+1)/2}B^{(7-p)/2}\frac{B'}B\right)'
+2r^{7-p}\left(A^{(p+1)/2}B^{(7-p)/2}\right)'
=-\frac{p+1}{8}\hspace{1pt}Q^2\hspace{1pt}\frac{e^{\frac{p-3}2\phi}A^{(p+1)/2}}{r^{8-p}B^{(7-p)/2}},
\label{abform}
\ee
and (from terms proportional to $x_ax_b$, after (\ref{uvform}) and (\ref{abform}) have been used to eliminate the terms depending on $Q$, i.e., originating from the form field)
\be
-\left(p\frac{A'}{A}+(8-p)\frac{B'}{B}\right)'+4\frac{A'}{A}\frac{B'}{B}+\frac{8-p}{r}\left(\frac{A'}{A}-\frac{B'}{B}\right)=\phi'^2.
\label{xaxbsmpl}
\ee
The $ua$-component of Einstein's equations gives
\be
-\left(p\frac{A'}{A}+(8-p)\frac{B'}{B}\right)^{\hspace{-1mm}\mbox{\bf .}}+4\frac{A'}{A}\frac{\dot{B}}{B}=\dot{\phi}\phi'.
\label{ua}
\ee
Finally, the $uu$-component of Einstein's equations (combined with the $uv$-component to eliminate the form) yields
\be
\begin{array}{l}
\dsty-\frac{A}{B}\,\frac{\left(r^{8-p}A^{(p+1)/2}B^{(7-p)/2}K'\right)'}{r^{8-p}A^{(p+1)/2}B^{(7-p)/2}}=\vspace{2mm}\\
\dsty\hspace{1.5cm}=(p-1)\left[\frac{\ddot{A}}{A}-\frac32\left(\frac{\dot{A}}{A}\right)^2\right]+(9-p)\left[\frac{\ddot{B}}{B}-\frac12\left(\frac{\dot{B}}{B}\right)^2-\frac{\dot{B}}{B}\frac{\dot{A}}{A}\right]+\dot{\phi}^2.
\end{array}
\label{K}
\ee

Equations (\ref{uvform}-\ref{xaxbsmpl}) are identical to those for the static ($u$-independent) problem, and should be solved first. Once that has been accomplished,
all the integration constants should be promoted to functions of $u$, and the result
should be substituted into (\ref{ua}), which will constrain the $u$-dependences.
Finally, (\ref{K}) will determine $K$. This algebraic structure essentially reduces
the $u$-dependent case to the $u$-independent one.

The solution for the $u$-independent case corresponding to our present ansatz has been given in \cite{liouville}. Essentially, one eliminates the $Q$-dependent terms (coming from the form field) from (\ref{dilatonform}) and (\ref{abform}) using (\ref{uvform})
to obtain
\begin{align}
&\left(r^{8-p}A^{(p+1)/2}B^{(7-p)/2}\left(\phi'-\frac{2(p-3)}{7-p}\frac{A'}A\right)\right)'=0,
\label{phi}\\
&\left(r^{15-2p}\left(A^{(p+1)/2}B^{(7-p)/2}\right)'\right)'=0.
\label{AB}
\end{align}
These equations are easily integrated, whereupon (\ref{uvform}) reduces to a Liouville equation (one-dimensional classical particle moving in an exponential potential) with respect to a new variable $\rho$ defined as $d/d\rho=r^{8-p}A^{(p+1)/2}B^{(7-p)/2}d/dr$. All the non-linearity of the problem becomes concentrated in this simple non-linear equation, which can be solved explicitly in terms of hyperbolic functions. Furthermore, as it turns out, equation (\ref{xaxbsmpl}) can be equivalently rewritten as an energy value specification for the above-mentioned Liouville equation and simply reduces to one constraint on the integration constants. We shall refer the reader to \cite{liouville} for explicit expressions.

Even though the static ($u$-independent) problem can be solved explicitly for our ansatz, it appears to be of limited use for general {\it non-extremal} $p$-branes.
The ansatz we have chosen was not the most general one allowed by the symmetries of the problem (though it will suffice for constructing the {\it extremal} solutions we are aiming at, and help us to keep the derivations reasonably compact), and in the presence of strong non-linearities, one should expect all types of motion permitted by the symmetry constraints to mix. In particular, one should relax the equality of $g_{uv}$ and $g_{\alpha\alpha}$ (some of related static non-extremal solutions have been constructed in \cite{blackened}, and a rather general analysis has been presented in \cite{generalnonextremal}), and add a non-zero $g_{ua}$. In our present investigations,
we shall not pursue this computation-extensive program, concentrating instead on the
case of extremal $p$-branes, which can be completely analyzed using the ansatz (\ref{metricansatz}-\ref{mformansatz}).

\section{Extremal solutions}

To obtain extremal $p$-brane solutions, we take particular integrals of (\ref{phi}) and (\ref{AB}), namely:
\be
\left(A^{(p+1)/2}B^{(7-p)/2}\right)'=0,\qquad
\phi'-\frac{2(p-3)}{7-p}\frac{A'}A=0.
\label{1storder}
\ee
(These particular integrals are known to correspond to extremal $p$-branes for the $u$-independent case.) One can then take
\be
A\propto \left(1+\frac{R^{7-p}}{r^{7-p}}\right)^{(p-7)/8}
\ee
(where $R$ will turn out to be simply another parametrization for the brane charge $Q$; we shall restore the expressions for the form field explicitly in our final results), compute the corresponding $B$ and $\phi$ using (\ref{1storder}), and check that the resulting $A$, $B$ and $\phi$ solve both (\ref{uvform}) and (\ref{xaxbsmpl}). Equations (\ref{uvform}-\ref{xaxbsmpl}) have now been satisfied.

As explained in the previous section, one needs to further promote all the integration
constants to functions of $u$ and solve (\ref{ua}) and (\ref{K}). The $u$-dependent prefactor in $g_{uv}$ can be changed arbitrarily by a redefinition of $u$, and we can use this freedom to relate the $u$-dependent prefactor of $A$ to the $u$-dependence of the dilaton. We thus introduce the following expressions to be substituted into (\ref{ua}) and (\ref{K}):
\be
\begin{array}{l}
\dsty A=e^{-f(u)/2}\left(1+h(u)\frac{R^{7-p}}{r^{7-p}}\right)^{(p-7)/8},\vspace{2mm}\\ \dsty B=\mu(u)e^{-f(u)/2}\left(1+h(u)\frac{R^{7-p}}{r^{7-p}}\right)^{(p+1)/8},\vspace{2mm}\\
\dsty \phi = f(u) +\frac{3-p}4\ln\left(1+h(u)\frac{R^{7-p}}{r^{7-p}}\right).
\end{array}
\label{extremalansatz}
\ee
This ansatz is designed to make the large $r$ asymptotics in {\it string frame} ($ds^2\equiv e^{\phi/2}ds_E^2$) look simple, as we choose to parametrize our solutions by this asymptotics. $g_{uv}$ is set to go to 1 for large $r$ (in string frame) as a matter of gauge choice; $g_{\alpha\alpha}$ is forced to go to 1 for large $r$ by hand. Equation (\ref{ua}) then yields
\be
\frac{\dot{h}}{h}=\dot{f}-\frac{7-p}2\frac{\dot{\mu}}{\mu},\qquad\qquad h=\frac{e^f}{\mu^{(7-p)/2}}
\ee
(the integration constant can always be absorbed into $R$). Equation (\ref{K}) becomes
\be
\frac{\left(r^{8-p}K'\right)'}{\mu r^{8-p}}=\left[4\ddot{f}-(9-p)\left(\frac{\ddot{\mu}}{\mu}-\frac{\dot{\mu}^2}{2\mu^2}\right)\right]+\frac{2 e^f R^{7-p}}{\left(\sqrt{\mu}r\right)^{7-p}}\left[\ddot{f}-\dot{f}\frac{\dot{\mu}}{\mu}-\frac{\ddot{\mu}}{\mu}+\frac{9-p}4\frac{\dot{\mu}^2}{\mu^2}\right],
\label{Kextr}
\ee
which is easily integrated to obtain a specific combination of $r^2$ and $1/r^{5-p}$ dependences\footnote{It is always possible to add terms solving the homogeneous version of (\ref{Kextr}), i.e., $r^0$ and $1/r^{7-p}$ with {\it arbitrary} $u$-dependent coefficients. The $r$-independent term can be absorbed into a redefinition of $v$. The $1/r^{7-p}$ term describes a peculiar singular pp-wave that propagates parallel to the brane essentially not interacting with it (in the sense that the shape of this wave does not affect the metric apart from its $uu$-component). We shall ignore these terms in our present considerations.\label{homogeneous}}.

If we now examine the large $r$ asymptotics of our solutions in {\it string frame}, we obtain:
\be
ds^2\equiv e^{\phi/2}ds_E^2=-2dudv+K(u,r)du^2+dy_\alpha^2+\mu(u)dx_a^2.
\ee
As indicated above, $K(u,r)$ contains an $r^2$ term, so the asymptotics indeed look like a plane wave. It is known, however, that, by redefining $v$ and $x^a$, plane wave metrics can always be put into a form that makes the $r^2du^2$ term in the metric vanish, with the wave profile encoded in $\mu(u)$ (the Rosen form), or into a form that makes $\mu(u)=1$, with the wave profile encoded in the coefficient of the $r^2du^2$ term in the metric (the Brinkmann form). Not surprisingly, this kind of transformations can be extended to our entire $p$-brane solutions (at all values of $r$).

More specifically, one can check that the transformation
\be
v=\tilde v + \mu(u)\eta(u)\dot{\eta}(u)\left(\frac{\tilde r^2}2+h(u)\left(\frac{R}{\eta(u)}\right)^{7-p}\frac{\tilde r^{p-5}}{p-5}\right),\qquad x^a=\eta(u)\tilde x^a
\label{brnkrsn}
\ee
preserves the algebraic form of our ansatz given by (\ref{metricansatz}) and (\ref{extremalansatz}), while multiplying $\mu$ by $\eta^2$. Since $\eta$ is an arbitrary function of $u$, it can be used to set $\mu$ to 1, in which case our $p$-brane solution is parametrized in a way that approaches the Brinkmann form of the plane wave in the asymptotic region. If this coordinate system is chosen, (\ref{Kextr}) simplifies further
and is integrated to yield
\be
K=\ddot{f}r^2\left(\frac{2}{9-p}-\frac{e^f}{5-p}\frac{R^{7-p}}{r^{7-p}}\right).
\ee
(Of course, other parametrization choices can be made, with (\ref{extremalansatz})-(\ref{Kextr}) giving the appropriate solutions; also, as already mentioned, we do not include the homogeneous solutions of (\ref{Kextr}) into our expressions.) With all the ingredients assembled together, the asymptotically Brinkmann form of our extremal plane-wave-$p$-brane solutions can be written, in string frame, as follows:
\be
\begin{array}{l}
\dsty ds^2\equiv e^{\phi/2}ds_E^2=\left(1+e^{f(u)}\frac{R^{7-p}}{r^{7-p}}\right)^{1/2}dx_a^2\vspace{2mm}\\
\dsty\hspace{5mm}+\left(1+e^{f(u)}\frac{R^{7-p}}{r^{7-p}}\right)^{-1/2}\left[-2dudv+\ddot{f}(u)\,r^2\left(\frac{2}{9-p}-\frac{e^{f(u)}}{5-p}\frac{R^{7-p}}{r^{7-p}}\right)du^2+dy_\alpha^2\right],\vspace{3mm}\\
\dsty \phi = f(u) +\frac{3-p}4\ln\left(1+e^{f(u)}\frac{R^{7-p}}{r^{7-p}}\right),\vspace{3mm}\\
\dsty F_{uv\alpha_1\cdots\alpha_{p-1}a}=\frac{x^a}{r}\,e^{-f(u)}\,\frac{\partial}{\partial r} \left(1+e^{f(u)}\frac{R^{7-p}}{r^{7-p}}\right)^{-1}\epsilon_{\alpha_1\cdots\alpha_{p-1}}\left[\frac{1}{\sqrt{2}}\right]_{p=3}\qquad (p\le 3),\vspace{3mm}\\
\dsty F_{a_1\cdots a_{8-p}}=\frac{x^a}{r}\,e^{-f(u)}\,\frac{\partial}{\partial r} \left(1+e^{f(u)}\frac{R^{7-p}}{r^{7-p}}\right)\epsilon_{a_1\cdots a_{8-p}a}\left[\frac{1}{\sqrt{2}}\right]_{p=3}\qquad (p\ge 3).
\end{array}
\label{solution}
\ee
For large values of $r$, this metric takes the form (ignoring the infrared problems for branes with a small number of transverse dimensions)
\be
ds^2=-2dudv+\frac{2}{9-p}\ddot{f}(u)\,r^2du^2+dy_\alpha^2+dx_a^2,\qquad \phi=f(u),
\label{asymptotics}
\ee
which is indeed the most general Brinkmann-coordinate plane wave (isotropic with respect to $x_a$-directions and with flat $y_\alpha$-directions), written in string frame.

Comparing our result to the previously published derivations, one can note that (22) of \cite{mbbsolns} becomes identical to our (\ref{solution}) for a specific choice of $f(u)$ in the dilaton profile as a linear function of $u$. For $p=1$, (21) of \cite{mbbsolns} corresponds\footnote{Incidentally, (39) of \cite{intersect2} presents a family of intersecting $p1$-$p5$-solutions that should reduce to (21) of \cite{mbbsolns} when the 5-brane charge is set to 0. \cite{intersect2} suggests that this family of solutions
should have two free parameters (three numbers, $a$, $b$ and $c$ with one quadratic constraint).
However, we believe that there is in fact only a one-parameter family in (39) of \cite{intersect2}, corresponding to the single parameter $Q$ of \cite{mbbsolns} (when the 5-brane charge is set to 0). An additional constraint on $a$, $b$ and $c$ of \cite{intersect2} (restoring the correspondence between (39) of \cite{intersect2} and (21) of \cite{mbbsolns}) can be derived by considering the $ur$-component of Einstein's equations.} to
a special choice of $f(u)$ in the dilaton profile (logarithmic in $u$, if the definition of $u$ is changed to agree with the one we are using), for which (in the asymptotically Rosen frame, different from the one used in (\ref{solution}) and related to it by transformations
of the form (\ref{brnkrsn})), the $du^2$ term disappears from the metric and the $u$- and $r$-dependences factorize throughout. For $p>1$, (21) of \cite{mbbsolns} corresponds to a plane wave asymptotics different from (\ref{asymptotics}), with non-trivial $y_\alpha$ polarizations present in the asymptotic plane wave (there is a $u$-dependent function
multiplying $dy_\alpha^2$ in the asymptotic expression for the metric). We have not considered such asymptotic plane waves here for the sake of compactness, but one should not expect any considerable complications in including them (the brane geometry is trivial
in the longitudinal directions, so superposing plane waves polarized in $y_\alpha$-directions
on it should be even simpler than for the case of $x_a$-directions). The reason why only special choices of the functional shape of the asymptotic plane wave appeared in the previous publications is that assumptions have been made about $u$- and $r$-dependence factorization, or about the absence of $du^2$ terms in the metric. By relaxing these assumptions, we have restored the functional arbitrariness of the asymptotic plane wave profile.

\section{Supersymmetry}

The fact that, in constructing our solutions, we have relied on the particular integrals (\ref{1storder}) of the equations of motion (which, for the $u$-independent case, correspond to extremal BPS solutions) makes it natural to expect that our $u$-dependent solutions will likewise be supersymmetric (and thus related to the D-branes of string theory). We shall now verify this proposition.

The supersymmetry transformations of the dilatino and the gravitino in string frame are given by \cite{bergshoeff, kalloshkumar}
\begin{equation}
 \delta \lambda =  (\partial_{\mu}\phi )\Gamma^{\mu}\varepsilon +\frac{3-p}{4 (p+2)!}e^{\phi} F_{\mu_{1}\dots \,  \mu_{p+2}}\Gamma^{\mu_{1}\dots \mu_{p+2}} \varepsilon_{(p)}' \, ,
\end{equation}
\begin{equation}
 \delta \psi_{\mu} =  \left( \partial_{\mu} + \frac{1}{4}\omega_{\mu\,\hat{\mu}\hat{\nu}}\gamma^{\hat{\mu}\hat{\nu}} \right)\varepsilon + \frac{(-1)^{p}}{8 (p+2)!} e^{\phi}F_{\mu_{1}\dots \mu_{p+2}}\Gamma^{\mu_{1}\dots  \mu_{p+2}} \Gamma_{\mu} \varepsilon_{(p)}' \, , 
\end{equation}
where $\gamma^{\hat{\mu}}$ are the Minkowski space $\gamma$-matrices and $\Gamma^{\mu} = e^{\mu}_{\hat{\mu}}\gamma^{\hat{\mu}}$, with $e_{\mu}^{\hat{\mu}}$ being the vielbein and the hatted indices referring to the tangent Minkowski space-time. $\varepsilon$ is a Majorana spinor for type IIA and a complex Weyl spinor for type IIB  and  $\varepsilon'$ is defined as:
\begin{equation}
\varepsilon'_{(p = 1,5)}  = i\varepsilon^{\star}, \qquad  \varepsilon'_{(p = 3)} = i\varepsilon,  \qquad   \varepsilon'_{ (p = 2,6)} = \ga_{11}\varepsilon,  \qquad \varepsilon'_{ (p = 4)}  = \varepsilon \, . 
\end{equation}
These supersymmetry variations are written in a formalism where both form fields and their duals are explicitly present, and we should use the duals of the forms of (\ref{metricansatz}-\ref{mformansatz}) for $p>3$ (we shall also not consider explicitly the $p=3$ self-dual case for the sake of compactness).

We shall examine the supersymmetry variations for the following ansatz, written in string frame:
\begin{equation}
ds^{2} = A_{s}(u,r) ( -2dudv + K(u,r) du^{2} +dy^{2}_{\alpha}) + B_{s}(u,r) dx_{a}^{2} \, ,
\label{anstz_start}
\end{equation}
with
\begin{align}
A_{s}&= e^{\phi/2} A =  \left(1+h(u)\frac{R^{7-p}}{r^{7-p}}\right)^{-1/2} \, , \\
B_{s}&= e^{\phi/2} B =\mu(u) \left(1+h(u)\frac{R^{7-p}}{r^{7-p}}\right)^{1/2} \,  
\end{align}
and 
\begin{align}
\phi& = f(u) + \frac{3-p}{4} \ln\left(1+h(u)\frac{R^{7-p}}{r^{7-p}}\right) = f(u) - \frac{3-p}{2} \ln A_{s}\, , \\
F_{uv\alpha_{1}.....\alpha_{p-1}a} &= e^{-f} \frac{x^{a}}{r} \partial_{r} \left(1+h(u)\frac{R^{7-p}}{r^{7-p}}\right)^{-1} \epsilon_{\alpha_{1}\dots \alpha_{p-1}}  = 2e^{-f} A_{s}' A_{s} \frac{x^{a}}{r}  \epsilon_{\alpha_{1}\dots \alpha_{p-1}}\, .
\label{anstz_end}
\end{align}
This ansatz includes (and is considerably more general than) our solution (\ref{solution}).

The  supersymmetry variations are given by
\be
\begin{array}{l}
\dsty\delta \lambda = \left( \dot f - \frac{3-p}{2}\frac{\dot A_{s}}{A_{s}} \right) A_{s}^{-1/2}\gamma^{\hat{u}}\varepsilon - \frac{3-p}{2} \frac{A_{s}'}{A_{s}B_{s}^{1/2}}\left[ \frac{x^{\hat{a}}}{r}\gamma^{\hat{a}}\varepsilon -\frac{ \epsilon_{\hat{\alpha}_{1}\dots\hat{\alpha}_{p-1}}}{(p-1)!}\gamma^{\hat{u}\hat{v} \hat{\alpha}_{1}\dots\hat{\alpha}_{p-1}\hat{a}}\frac{x^{\hat{a}}}{r}   \varepsilon' \right] \, ,\vspace{2mm}\\  
\dsty \delta \psi_{u}  =  \left( \partial_{u} - \frac{1}{4}\frac{\dot A_{s}}{A_{s}}\gamma^{\hat{v}\hat{u}} + \frac{1}{4}\frac{ A_{s}^{1/2}}{B_{s}^{1/2}} K'\gamma^{\hat{u}\hat{a}}\frac{x^{\hat{a}}}{r}  \right)\varepsilon \vspace{2mm} \\
\dsty\hspace{2cm}- \frac{1}{4}\frac{A_{s}'}{A_{s}^{1/2}B_{s}^{1/2}}\left(\gamma^{\hat{v}} - \frac{K}{2}\gamma^{\hat{u}} \right) \left[ \frac{x^{\hat{a}}}{r}\gamma^{\hat{a}}\varepsilon -\frac{ \epsilon_{\hat{\alpha}_{1}\dots\hat{\alpha}_{p-1}}}{(p-1)!}\gamma^{\hat{u}\hat{v} \hat{\alpha}_{1}\dots\hat{\alpha}_{p-1}\hat{a}}\frac{x^{\hat{a}}}{r} \varepsilon' \right] \, , \vspace{2mm}\\
\dsty \delta \psi_{v}  =   \partial_{v} \varepsilon - \frac{1}{4}\frac{A_{s}'}{A_{s}^{1/2}B_{s}^{1/2}}\gamma^{\hat{u}} \left[ \frac{x^{\hat{a}}}{r}\gamma^{\hat{a}}\varepsilon -\frac{ \epsilon_{\hat{\alpha}_{1}\dots\hat{\alpha}_{p-1}}}{(p-1)!}\gamma^{\hat{u}\hat{v} \hat{\alpha}_{1}\dots\hat{\alpha}_{p-1}\hat{a}}\frac{x^{\hat{a}}}{r} \varepsilon' \right] \, , \vspace{2mm} \\
\dsty \delta \psi_{\alpha}  =   \left( \partial_{\alpha} -   \frac{1}{4}\frac{\dot A_{s}}{A_{s}} \gamma^{\hat{u} \hat{\alpha}} \right) \varepsilon + \frac{1}{4}\frac{A_{s}'}{A_{s}^{1/2}B_{s}^{1/2}}\gamma^{\hat{\alpha}} \left[ \frac{x^{\hat{a}}}{r}\gamma^{\hat{a}}\varepsilon -\frac{ \epsilon_{\hat{\alpha}_{1}\dots\hat{\alpha}_{p-1}}}{(p-1)!}\gamma^{\hat{u}\hat{v} \hat{\alpha}_{1}\dots\hat{\alpha}_{p-1}\hat{a}}\frac{x^{\hat{a}}}{r} \varepsilon' \right] \, , \vspace{2mm} \\
\dsty \delta \psi_{a}  =   \left( \partial_{a} -   \frac{1}{4}\frac{\dot B_{s}}{ A_{s}^{1/2} B_{s}^{1/2} } \gamma^{\hat{u} \hat{a}} \right) \varepsilon - \frac{1}{4}\frac{A_{s}'}{A_{s}} \gamma^{\hat{a}}  \sum_{\hat{b}\neq \hat{a}} \left[ \frac{x^{\hat{b}}}{r}\gamma^{\hat{b}}\varepsilon -\frac{ \epsilon_{\hat{\alpha}_{1}\dots\hat{\alpha}_{p-1}}}{(p-1)!}\gamma^{\hat{u}\hat{v} \hat{\alpha}_{1}\dots\hat{\alpha}_{p-1}\hat{b}}\frac{x^{\hat{b}}}{r} \varepsilon'  \right]  \vspace{2mm} \\
 \dsty\hspace{4cm} + \frac{(-1)^p}{4}\frac{A_{s}'}{A_{s}} \frac{ \epsilon_{\hat{\alpha}_{1}\dots\hat{\alpha}_{p-1}}}{(p-1)!}\gamma^{\hat{u}\hat{v} \hat{\alpha}_{1}\dots\hat{\alpha}_{p-1}}\frac{x^a}{r} \varepsilon' \, 
\end{array}
\ee
(where $x^{\hat{a}}$ simply denotes $x^a$ with the numerical value of $a$ set equal to $\hat{a}$, and summation over repeated indices is understood). These all vanish if 
\begin{equation}
\varepsilon = A_{s}^{1/4} \tilde\varepsilon\, , 
\end{equation}
where $\tilde\varepsilon$ is a constant spinor such that
\begin{equation}
\gamma^{\hat{u}} \tilde\varepsilon = 0 \,  ,
\end{equation}
and 
\begin{equation}
\gamma^{\hat{a}} \tilde\varepsilon -\frac{ \epsilon_{\hat{\alpha}_{1}\dots\hat{\alpha}_{p-1}}}{(p-1)!}\gamma^{\hat{u}\hat{v} \hat{\alpha}_{1}\dots\hat{\alpha}_{p-1}\hat{a}}\tilde\varepsilon' =0 \, ,
\end{equation}
with $\tilde\varepsilon' $  defined similarly to $\varepsilon'$, which makes 8 supersymmetries manifest for our solutions and establishes them as the BPS $p$-branes. Note that the presence of these supersymmetries is insensitive to whether the equations of motion are satisfied, as long as the field configuration is of the form (\ref{anstz_start}-\ref{anstz_end}).

\section{Conclusions}

We have presented a family of ten-dimensional supergravity solutions describing extended extremal $p$-branes embedded into a dilaton-gravity plane wave, with the brane worldvolume aligned along the propagation direction of the wave. We have assumed an isotropic plane wave polarization in the directions transverse to the brane worldvolume, and the absence
of polarization components along the brane worldvolume. No assumptions have been made
about the functional shape of the plane wave profile, which is contained in our family of solutions as an arbitrary function of the light-cone time.

It could be very interesting and important to generalize our results to the case of 0-branes. In that case, there is no worldvolume to be aligned with the propagation direction of the wave, and the 0-brane is subject to forces induced by the plane wave.
However, it can be seen from the corresponding D0-brane DBI analysis that
there are configurations for which the gravity and dilaton forces balance each other and the 0-brane does not move. One could expect relatively simple supergravity solutions for these cases, and they are also precisely the solutions whose near-horizon geometry may have significance\footnote{We thank S.~Minwalla for
suggesting this.} within the context of time-dependent matrix models. Unfortunately,
our present derivations do not allow to construct such solutions.

After this work had been completed, a recent preprint \cite{recent}
addressing very similar issues came to our attention.
In that publication, a somewhat more general ansatz
(compared to the one we have used here) is examined (non-trivial
asymptotic plane wave polarizations in the directions parallel
to the brane are added); considerations are also given
to intersecting brane solutions. The advantage of our
present treatment is that all the light-cone time dependences
are derived explicitly (in \cite{recent}, the problem is reduced to
ordinary differential equations, which are not solved),
the equation determining the $uu$-component of the metric (\ref{metricansatz})
is analyzed without any assumptions (this analysis does not confirm\footnote{(2.42) of \cite{recent} assumes that $K$ of (\ref{metricansatz}) is a combination of $r^0$ and $1/r^{7-p}$ dependences on $r$. As is evident from our analysis in section 3, however, an inclusion of $r^2$ and $1/r^{5-p}$ dependences is essential for maintaining the functional arbitrariness of the plane wave profile.
The inclusion of $r^0$ and $1/r^{7-p}$ terms is optional, as far as the construction of plane-wave-$p$-brane solutions is concerned, cf. footnote \ref{homogeneous}.} the suggestions of \cite{recent}), and possible generalizations
to the non-extremal case are contemplated.

\section{Acknowledgments}
 
 This research has been
supported in part by the Belgian Federal Science Policy Office through the Interuniversity Attraction Pole IAP VI/11 and by FWO-Vlaanderen through project G.0428.06. The work
of F.G. is supported by a Ph.D. fellowship from the International Solvay
Institutes.

\appendix

\section{The equations of motion}

For the reader's convenience, we present here an explicit unprocessed form of the equations of motion (\ref{Einstein}-\ref{bianchi}) for our ansatz (\ref{metricansatz}-\ref{mformansatz}). The $uu$-component of Einstein's equations (\ref{Einstein}) reads
\be
\begin{array}{l}
\dsty-\frac{p-1}2\left[\frac{\ddot{A}}{A}-\frac32\left(\frac{\dot{A}}{A}\right)^2\right]-\frac{9-p}2\left[\frac{\ddot{B}}{B}-\frac12\left(\frac{\dot{B}}{B}\right)^2-\frac{\dot{B}}{B}\frac{\dot{A}}{A}\right]+\frac{K'A'}{2B}\vspace{2mm}\\
\dsty-\frac1{r^{8-p}}\left(r^{8-p}\frac{(KA)'}{2B}\right)'-\frac{(KA)'}{2B}\left(\frac{p-1}{2}\frac{A'}{A}+\frac{9-p}2\frac{B'}{B}\right)=\frac12\dot{\phi}^2+\frac{p-7}{16}e^{\frac{p-3}2\phi}\frac{KF^2A}{B^{8-p}}.
\end{array}
\ee

\noindent The $uv$-component (identical to the $\alpha\alpha$-components for our ansatz):
\be
\left(\frac{A'}{2B}\right)'+\frac{8-p}r\left(\frac{A'}{2B}\right)+\frac{A'}{2B}\left(\frac{p-1}{2}\frac{A'}{A}+\frac{9-p}2\frac{B'}{B}\right)=\frac{7-p}{16}e^{\frac{p-3}2\phi}\frac{F^2A}{B^{8-p}}.
\ee

\noindent The $ua$-component:
\be
-\left(\frac{p}{2}\frac{\dot{A}}{A}+\frac{8-p}2\frac{\dot{B}}{B}\right)'+2\frac{\dot{B}}{B}\frac{A'}{A}=\frac12\dot{\phi}\phi'.
\ee

\noindent The $ab$-component (terms proportional to $\delta_{ab}$):
\be
-\left(\frac{B'}{2B}\right)'+\frac{2p-15}{2r}\frac{B'}{B}-\frac{p+1}{2r}\frac{A'}{A}-\frac{p+1}4\frac{A'}{A}\frac{B'}B+\frac{p-7}4\left(\frac{B'}{B}\right)^2=\frac{p+1}{16}e^{\frac{p-3}2\phi}\frac{F^2}{B^{7-p}}.
\ee

\noindent The $ab$-component (terms proportional to $x_ax_b$):
\be
\begin{array}{l}
\dsty (p-7)\left[\left(\frac{B'}{2B}\right)'-\frac1{2r}\frac{B'}{B}\right]-(p+1)\left[\left(\frac{A'}{2A}\right)'-\frac1{2r}\frac{A'}{A}\right]\vspace{2mm}\\
\dsty\hspace{2cm}-\frac{p+1}{4}\frac{A'^2}{A^2}+\frac{p+1}2\frac{A'}A\frac{B'}B+\frac{7-p}4\frac{B'^2}{B^2}=\frac12\phi'^2-e^{\frac{p-3}2\phi}\frac{F^2}{2B^{7-p}}.
\end{array}
\ee

\noindent The remaining components are identically zero.
The dilaton equation (\ref{dila}) can be written as
\be
\frac{\left(r^{8-p}A^{(p+1)/2}B^{(7-p)/2}\phi'\right)'}{r^{8-p}A^{(p+1)/2}B^{(9-p)/2}}=\frac{p-3}4e^{\frac{p-3}2\phi}\frac{F^2}{B^{8-p}}.
\ee

\noindent Finally, the equations for the form (\ref{field}-\ref{bianchi}) simply yield:
\be
\left(r^{(8-p)}F\right)'=0,\qquad \dot{F}=0.
\ee


\end{document}